\documentclass[12pt,preprint]{aastex}
\usepackage{natbib,hyperref,verbatim,CJK}
\begin{document}

\title{Searching for Be Stars in the Open Cluster NGC\,663}

\begin{CJK*}{UTF8}{bkai}

\author{
P.~C.~Yu (俞伯傑)\altaffilmark{1},
C.~C.~Lin (林建爭)\altaffilmark{1},
W.~P.~Chen (陳文屏)\altaffilmark{1,2},
C.~D.~Lee (李建德)\altaffilmark{1},
W.~H.~Ip (葉永烜)\altaffilmark{1,3},
C.~C.~Ngeow (饒兆聰)\altaffilmark{1},
Russ Laher (良主嶺亞)\altaffilmark{4},
Jason Surace\altaffilmark{4}, \&
Shrinivas R. Kulkarni\altaffilmark{5}
}

%=========%=========%=========%=========%=========%=========%=========%=========
\altaffiltext{1}{Graduate Institute of Astronomy, National Central University, 300 Jhongda Road, Jhongli 32001, Taiwan}
\altaffiltext{2}{Department of Physics, National Central University, 300 Jhongda Road, Jhongli 32001, Taiwan}
\altaffiltext{3}{Institute of Space Science, National Central University, 300 Jhongda Road, Jhongli 32001, Taiwan}
\altaffiltext{4}{Spitzer Science Center, California Institute of Technology, M/S 314-6, Pasadena, CA 91125, USA} 
\altaffiltext{5}{Division of Physics, Mathematics and Astronomy, California Institute of Technology, Pasadena, CA 91125, USA} 
%=========%=========%=========%=========%=========%=========%=========%=========

\begin{abstract}
We present Be star candidates in the open cluster NGC\,663, identified by H$\alpha$ imaging photometry with the Palomar Transient Factory Survey, 
as a pilot program to investigate how the Be star phenomena, the emission spectra, extended circumstellar envelopes, and fast rotation, correlate with massive stellar evolution.  
Stellar membership of the candidates was verified by 2MASS magnitudes and colors, and by proper motions. 
We discover 4 new Be stars and exclude one known Be star from being a member due to its inconsistent proper motions. 
The fraction of Be stars to member stars [N(Be)/N(members)] in NGC\,663 is 3.5\%. 
The spectral type of the 34 Be stars in NGC\,663 shows bimodal peaks at B0--B2 and B5--B7, which is consistent with the statistics in most star clusters. 
Additionally, we also discover 23 emission-line stars of different types, including non-member Be stars, dwarfs, and giants.

\end{abstract}

\keywords{galaxies: star clusters: individual (NGC~663) -- stars: emission-line -- stars: Be}

\section{Introduction}

Emission-line stars are symbolized by Balmer lines, particularly the H$\alpha$ lines in emission spectra.
In general, the mechanisms responsible for the emission lines include mass accretion and chromospheric activity.
Emission stars can be in almost any stage during stellar evolution, from pre-main
sequence T Tauri and Herbig Ae/Be stars, main-sequence Be/Ae and dMe stars, Wolf-Rayet stars, to giants and
supergiants.  Among these, the classical Be stars are particularly interesting because, other than the emission spectra,
they also show near-infrared-excess emission above the stellar photospheric radiation, and fast rotation with
an equatorial speed up to $70 - 80\%$ of breakup velocity \citep{tow04}.

The infrared excess of Be stars is usually attributed to plasma free-free emission in the extended envelopes.
Even though it is thought that the fast rotation must play a decisive role, the interplay of the Be phenomena,
namely the emission lines, infrared excess, and rapid rotation, is still unclear.  One school of thought
proposes that classical Be stars become fast rotators only at the second half of the main sequence \citep{fab00},
or near the turn-off \citep{kel01}. These Be stars are found in young ($\sim 10$~Myr)
star clusters in the Milky Way galaxy \citep{mat08}, LMC, and SMC \citep{wis06}. Alternatively, the stars
might have been spun up by mass transfer in binaries \citep{mcs05}.

Care should be exercised when collecting the sample of classical Be stars in a young cluster because a distinctly different group, the
Herbig Ae/Be stars, intermediate-mass pre-main sequence objects, exhibit very similar observational properties,
such as the emission spectra and infrared excess. \citet{lee11} analyzed a group of Be stars away from
any signposts of recent star formation, thereby being excluded of the pre-main sequence status, yet have
unusually large infrared excess, with the observed 2MASS colors $J-H$ and $H-K_{s}$ both greater than 0.6~mag,
that must be accounted for by thermal dust emission.

The primary uncertainty in the age determination of a single Be star is rectified if Be stars in a sample of
star clusters at different ages can be studied. However, the sample of classical Be stars in open clusters
is not complete due to several reasons.  First, a nearby open cluster occupies a wide sky area, so a
comprehensive survey for emission-line stars is time consuming, and often limited to bright stars.  Secondly,
Be stars show photometric and spectroscopic variability.  With the wide field coverage of the Palomar Transient
Factory (PTF) project, we have initiated a program to search for emission stars in clusters
of different ages. This paper reports our methodology to identify emission-line stars with
the PTF H$\alpha$ images. 

Here we present the results for the first target, the young open cluster NGC\,663
($\alpha_{2000}=01^{\rm h}46^{\rm m}12^{\rm s}, \delta_{2000}=61\degr13\arcmin30\arcsec$; galactic coordinates $l, b$ = 129.470$\degr$, $-$0.953$\degr$).  
With an age of 31~Mys and at a distance 2.1~kpc \citep{kha13}, this cluster is known to host a large number of Be stars
\citep*{Sanduleak79,pig01,mathew11}. NGC\,663 was also considered as a part of stellar association Cas\,OB8. 
Cas\,OB8 has been reported to be located at a distance of 2.9 kpc \citep{dambis2001}, and thus NGC\,663 appears to be in the foreground. 
However, the distance of Cas\,OB8 seems to be overestimated. Using old and new reduction of the Hipparcos data, 
the mean distance of Cas\,OB8 is suggested to be 2.3 kpc \citep{melnik2009}. Taking possible uncertainty  of distance estimation into consideration,
NGC\,663 could be a part of Cas\,OB8.

In Section~2, we describe the acquisition of the observations and the analysis to
recognize emission stars.  In Section~3, we present our list of H$\alpha$ stars and compare our results,
which reach fainter magnitudes, with those in the literature.  Section~4 gives a summary of this study.

\section{Observations and Data Analysis}

Data used in this study include the PTF H$\alpha$ images to identify, and the 2MASS near-infrared photometry and
proper motions (PMs) to characterize, the emission star candidates. The PTF is an automated, wide-field survey
with a 7.3 square-deg field of view \citep{Law2009}.  All observations were carried out with the 48-inch Samuel
Oschin Telescope at Palomar Observatory. The H$\alpha$ and the continuum images are taken through HA656 and
HA663 narrow-band filters (hereafter H$\alpha$ and r-band, respectively), which have a central wavelength of 6563~\AA~
and 6630~\AA, with a bandwidth of 75~\AA. The exposure time for the H$\alpha$ and r-band images is 60 seconds.
The image data were processed for bias corrections, flat fielding, and astrometric calibration with pipelines developed by 
the Infrared Processing and Analysis Center \citep[IPAC;][]{Laher2014}.

The Two Micron All Sky Survey point source catalog \citep[2MASS;][]{cut03} provides a uniformly calibrated
database of the entire sky in the near-infrared $J$, $H$, and $K_{s}$ bands with a 10$\sigma$ detection limit of
15.8, 15.1, and 14.3~mag, respectively.  Kinematic information can be obtained to secure the membership in a
star cluster by the PPMXL data set, which is derived from an all-sky merged catalog based on the USNO-B1
and 2MASS positions of 900 million stars and galaxies, reaching a limiting V$\sim20$~mag \citep{ros10}.
The typical error of PMs is less than 2~milliarcseconds (mas) per year for the brightest stars
with Tycho-2 \citep{hog00} observations, and is more than 10~mas~yr$^{-1}$ at the faint limit.
However,  \citet{ros10} noted that about 90 million (10\%) objects of PPMXL include spurious entries; 
they found double or several matches of USNO-B1 stars with a 2MASS star. 
\citet{kha12} have averaged such PMs and computed their errors in sky areas with star clusters including NGC\,663. 
Therefore, we use these PMs instead of PPMXL data. 

In our study, we made use of the 2MASS photometry and the PMs to select and
characterize stellar member candidates.  While matching counterparts in different star catalogs, a coincidence radius
of 2\arcsec\ was used among PTF, 2MASS, and PMs sources.

\subsection{Data Analysis}

\subsubsection{The Search Region}

\begin{figure}
\epsscale{1}
\plotone{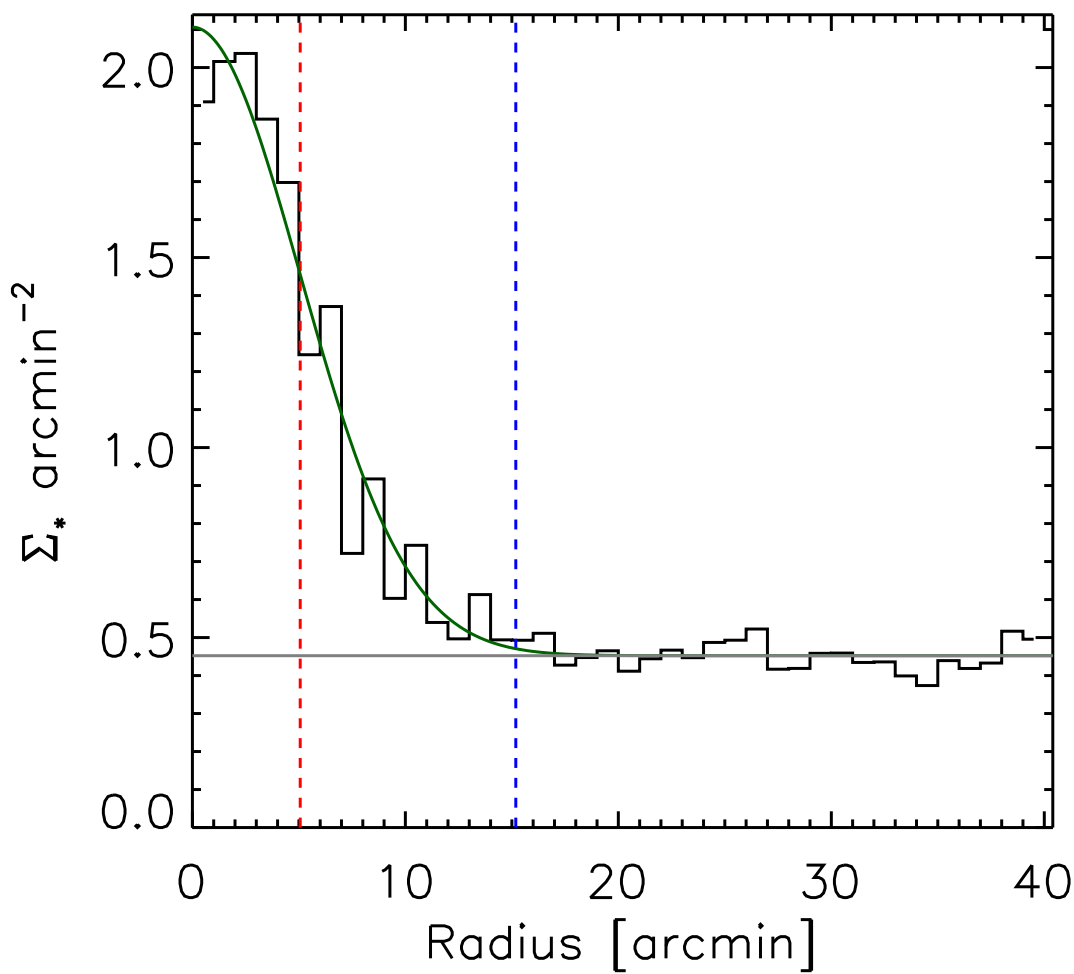}
\caption{ Radial density profile for NGC\,663. The grey horizontal line depicts the background density.  The green curve shows the best Gaussian fitting results. 
Red and blue vertical lines represent the width of the curve with $1\sigma$ and $3\sigma$, respectively. We adopted a box with the side of
40\arcmin~(4$\sigma$) as the field of NGC\,663 to cover the region of previous studies and possible candidates.}
\label{rdp}
\end{figure}

Open clusters generally have irregular shapes during their evolution either the internal relaxation process or the Galactic external
perturbation \citep{chen2004}. We therefore need to adopt a suitable region to include all possible member stars for NGC\,663. 
To determine the survey region, we selected the 2MASS point sources with $\mbox{S/N}\ge10$ in all $J, H, K_{s}$ bands and
with PMs uncertainties $< 5.0$~mas~yr$^{-1}$.  Based on the radial density profile (Figure~\ref{rdp}),
the best half-Gaussian fitting gives a 3$\sigma$ diameter of $\sim30\arcmin$, which corresponds to a diameter of $\sim$18~pc
at 2.1~kpc. Hence we adopted a box with the side of 40\arcmin~(4$\sigma$) as the field of NGC\,663 in the subsequent analysis to 
cover the region of previous studies and possible candidates.

\subsubsection{H$\alpha$ Photometry and Emission-Line Candidates}

\begin{figure*}
\epsscale{1.0}
\plotone{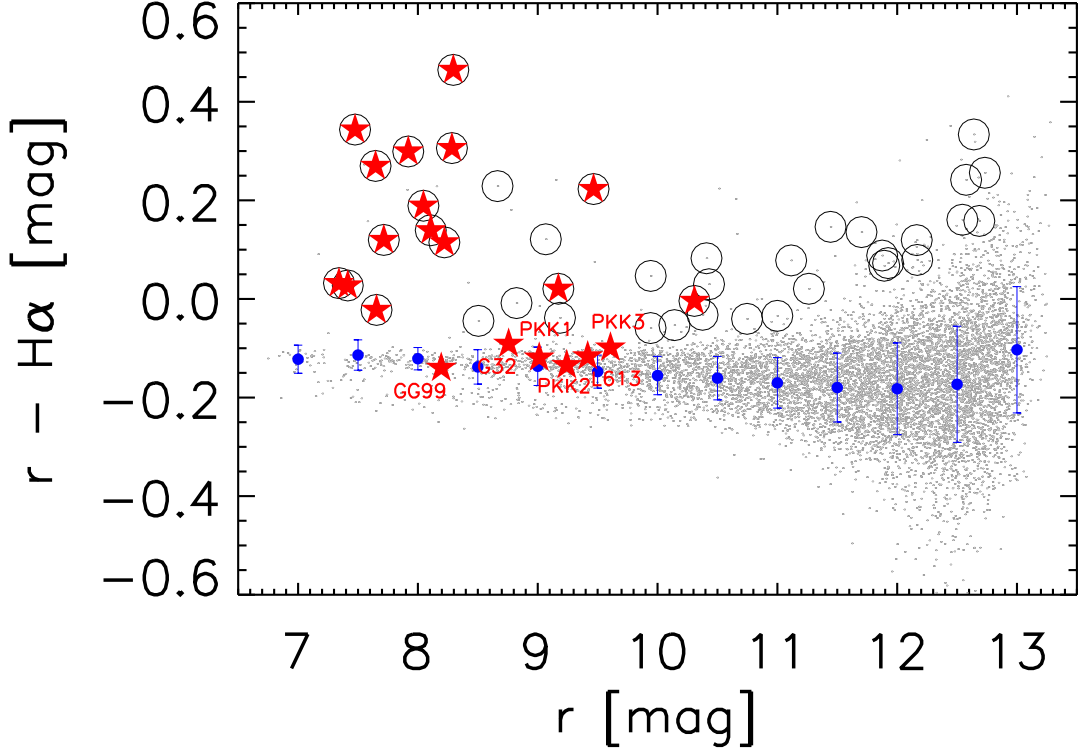}
\caption{ H$\alpha$ photometry. Grey dots: all stars within a field of ~2\degr~$\times$~1\degr; open circles: emission-line
candidates within a field of ~40\arcmin~$\times$~40\arcmin; red pentagrams: known Be stars in NGC\,663. Blue filled circles represent mean values of
r$-$H$\alpha$ values within each 0.5~mag r-band bin, and the photometric scattering are calculated with the error propagated from photometric and systematic errors.
The photometric error is the weighted photometric error of r$-$H$\alpha$ values for stars within the 0.5~mag bin 
while the systematic errors are the standard deviation of r$-$H$\alpha$ values for stars within the same bin.  
PKK1, PKK2, PKK3, GG99, G32, and L613 are known Be stars that are not selected by our method.}
\label{Ha}
\end{figure*}

\begin{figure}
\epsscale{1}
\plotone{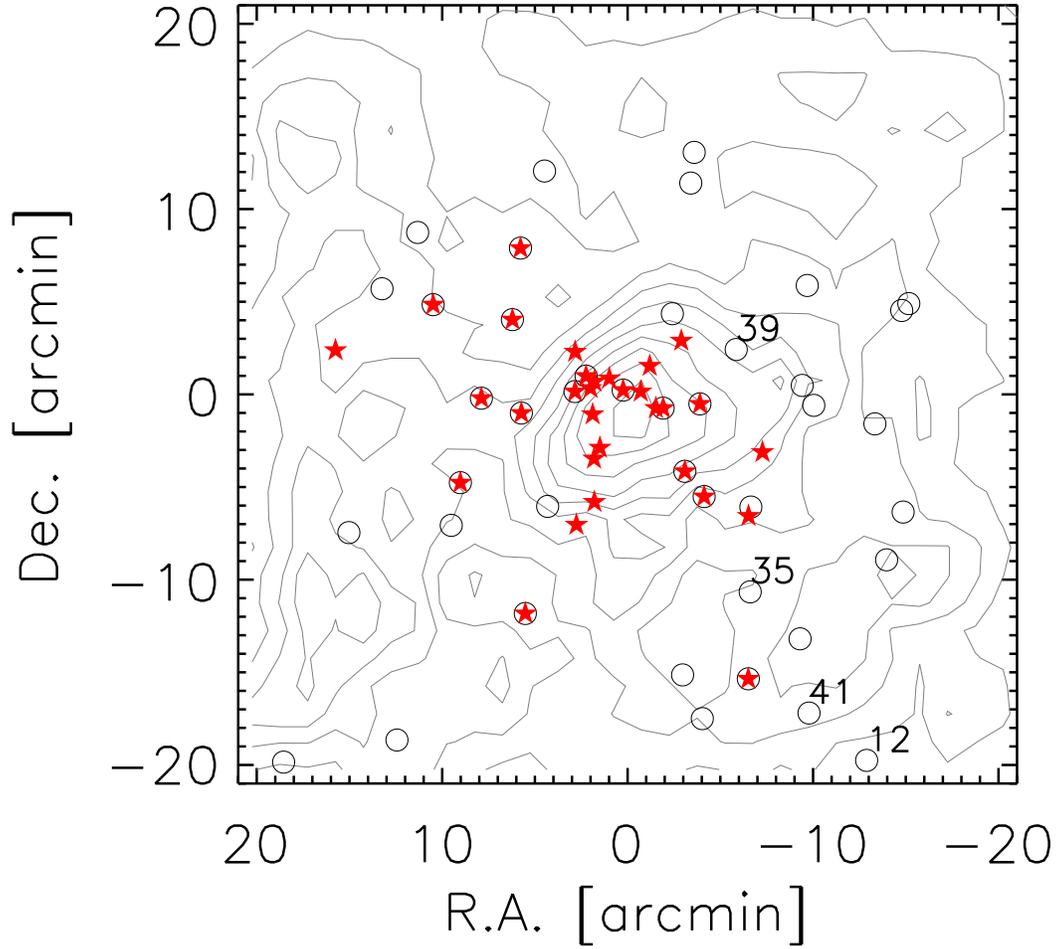}
\caption{Spatial distribution of known Be stars and emission-line candidates. Symbols are same as Figure~\ref{Ha}. The symbols with numbers represent ID numbers of 4 newly discovered Be stars.
Contours represent distribution of member stars within a field of ~40\arcmin~$\times$~40\arcmin.}
\label{spatial}
\end{figure}

To identify possible H$\alpha$ emission stars, we compute the difference of the instrumental magnitude
between the H$\alpha$ and the r-band images (e.g., r$-$H$\alpha$).  Under the assumption that the majority of stars
in the field exhibit neither H$\alpha$ in emission nor in absorption, those stand out in r$-$H$\alpha$ are
probable H$\alpha$ stars.  Figure~\ref{Ha} shows the r$-$H$\alpha$ values for all stars (grey dots) within a field of
$\sim 2\degr \times 1\degr$, which guarantees to cover the spatial extent of NGC\,663.
Because of increasing scattering of the r$-$H$\alpha$ values toward faint magnitudes due to photometric
errors, we exercised different selection criteria according to the brightness of the stars.
We first calculated the photometric scattering ($\sigma_{p}$) for each 0.5~mag r-band bin with the error propagated from photometric and
systematic errors. The photometric error is the weighted photometric error of r$-$H$\alpha$ values for stars within the 0.5~mag bin 
while the systematic errors are the standard deviation of r$-$H$\alpha$ values for stars within the same bin. 
Then we selected emission-line candidates for those with r$-$H$\alpha$ $>$ 2$\sigma_{p}$ in each 0.5~mag bin,
i.e., with significant flux excess in the H$\alpha$ image than in the $r$-band image, as illustrated in Figure~\ref{Ha}.
A total of 42 emission-line candidates were thus identified within a $40\arcmin \times 40\arcmin$ field.
Table~1 lists the properties of the 42 emission-line candidates. The first column provides identification numbers. Columns 2 and 3 provide coordinates in degree. Columns 4 and 5 show the PM measurements in right ascension and declination. 
Column 6 gives PM uncertainties. 
Subsequent columns, 7 to 9, list the 2MASS $K_{s}$, $J-H$, and $H-K_{s}$ values. Column 10 lists names of known Be stars. Columns 11 and 12 give the PTF instrumental r-band and r$-$H$\alpha$ magnitudes. Column 13 provides the object classifications adopted from SIMBAD. Figure~\ref{spatial} shows the spatial distribution of our candidates. 

\subsubsection{Identification of Membership}

\begin{figure}
\epsscale{1}
\plotone{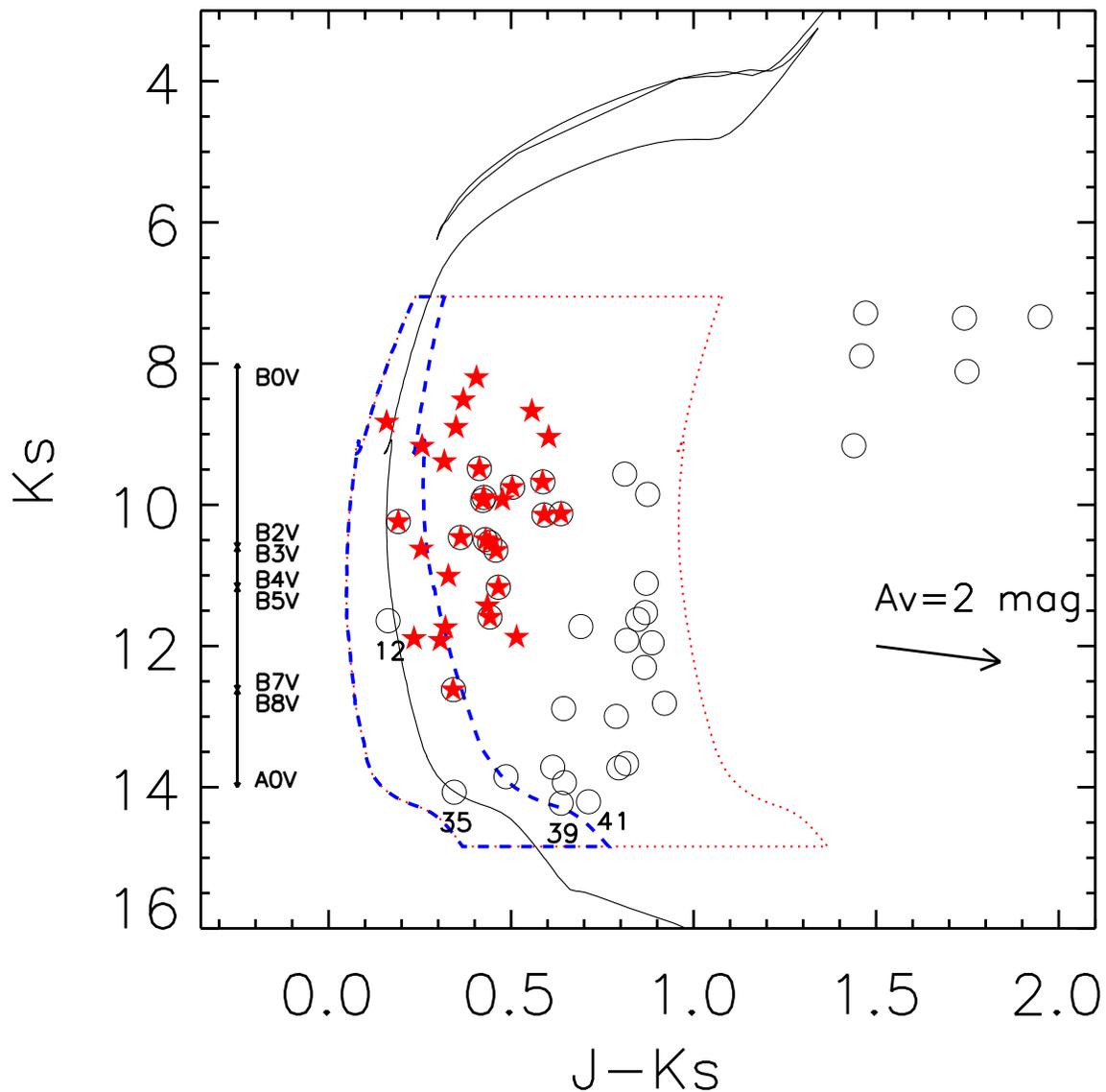}
\caption{2MASS color--magnitude diagram in the cluster region. The black solid line delineates the post-main-sequence isochrone \citep{gir02} for 31~Myr
located at 2.1~kpc \citep{kha13} with an interstellar extinction of $A_V\sim2.1$~mag. Symbols are same as Figure~\ref{Ha}.
Non-emission-line stars within the blue-dashed region are considered as photometric membership.
Emission-line candidates are considered as photometric membership with a wider red-dashed region.}
\label{cmd}
\end{figure}

We verify member stars of NGC\,663 with photometric and kinematic memberships.
First, we determine the photometric membership by selecting stars within the region that are near the isochrone \citep{gir02} in the $K_{s}$ versus $J-K_{s}$ color-magnitude diagram (CMD).
\citet{pandey05} suggest a range of age from 10 to 50~Myr for NGC\,663. Here we adopt the mean age of $\sim$ 31~Myr, the distance of 2.1~kpc, and E(B$-$V) of 0.7~mag
from \citet{kha13} for the isochrone. The region for the photometric membership is defined by estimating $J-K_{s}$ errors propagated from photometric errors of $J$ and $K_{s}$ values along 
with the isochrone from $K_{s}$ = 7~mag toward faint $K_{s}$ magnitudes (Figure~\ref{cmd}).
On the other hand, verification of the photometric membership for the H$\alpha$  emission-line candidates is not intuitive. \citet{lee11} showed that Be stars can have unusually large 
infrared excess with $J-H$ and $H-K_{s}$ both greater than 0.6 mag. As shown in Figure~\ref{cmd}, we thus extend the selection region of the photometric membership with $J-K_{s}~\sim$~1.2 mag for the emission-line candidates.

\begin{figure}
\epsscale{1}
\plotone{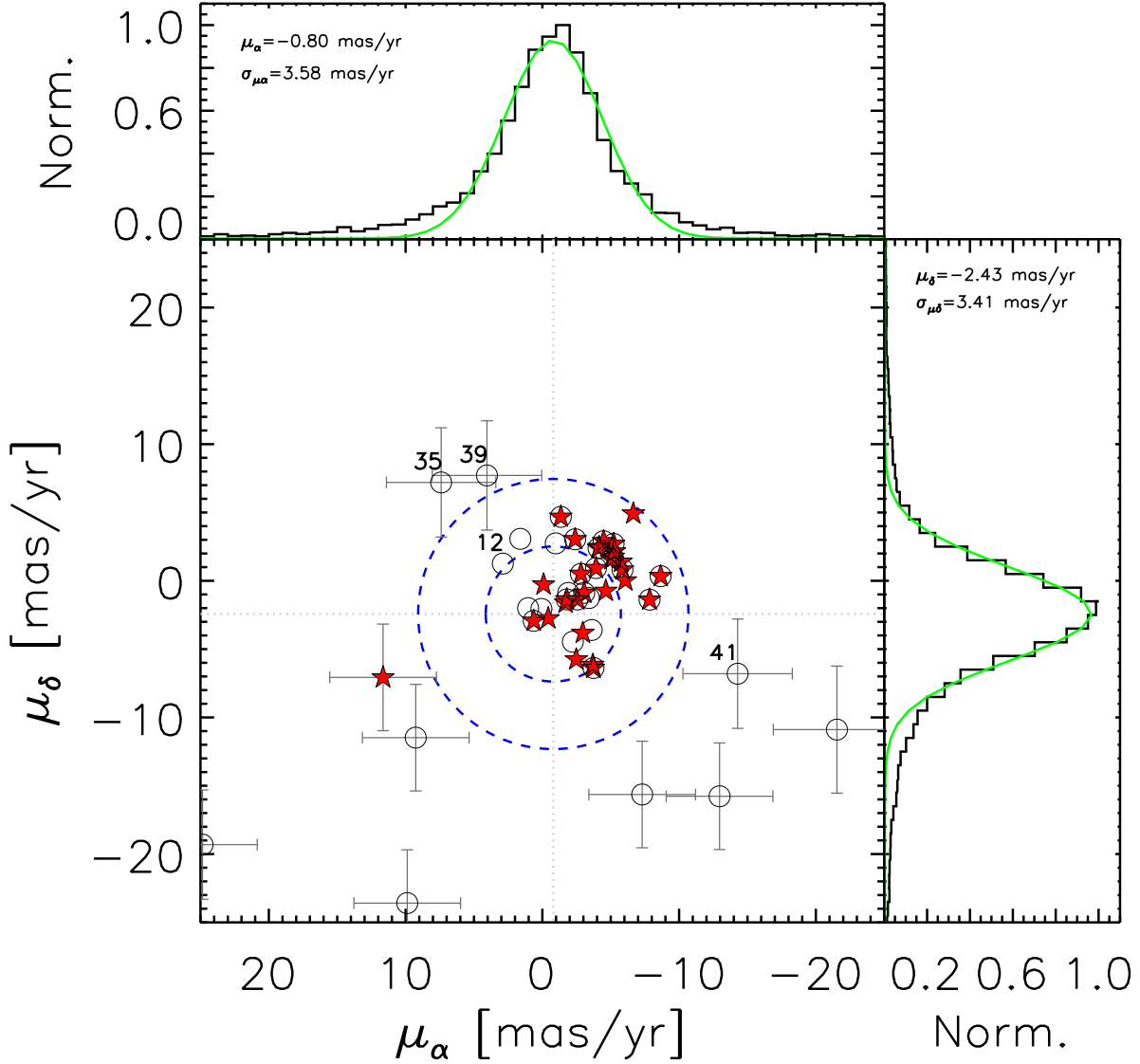}
\caption{Stellar PMs of known Be stars and emission-line candidates toward NGC\,663. The PM of NGC\,663 is ($\mu_{\alpha}$,$\mu_{\delta}$)=($-$0.80, $-$2.43)~mas~yr$^{-1}$.
The inner and outer blue-dashed circle are 1$\sigma_{\mu}$ and 2$\sigma_{\mu}$ regions  to define the probable PMs membership. 
Symbols are same as Figure~\ref{Ha}. The numbers represent the ID numbers of emission-line candidates listed in Table~1. 
The bars represent errors of PMs. Upper and right panels represent $\sigma_{\mu\alpha}$ and $\sigma_{\mu\delta}$ distribution of stars within a 40$\arcmin$ $\times$ 40$\arcmin$; green lines show gaussian fitting to PMs.}
\label{ppm}
\end{figure}

Secondly, we define a region that could include the possible kinematic membership.
We have calculated the averaged PM, and standard deviations of stars within the $40\arcmin \times 40\arcmin$ field by fitting a Gaussian distribution to PMs.
The mean PM is ($-$0.80, $-$2.43) mas/yr, and standard deviations $\sigma_{\mu\alpha}$ and $\sigma_{\mu\delta}$ are 3.58~mas~yr$^{-1}$ and 3.41~mas~yr$^{-1}$, respectively. 
We then adopt 2-sigma approach to determine the kinematic membership. 
The sigma $\sigma_{\mu}$ = 4.94 mas/yr is the error propagated from  $\sigma_{\mu\alpha}$ and $\sigma_{\mu\delta}$.
Stars are considered as the kinematic membership if they are located inside the 2$\sigma_{\mu}$ region (e.g., 2$\times$$\sigma_{\mu}$ = 9.88~mas~yr$^{-1}$) (Figure~\ref{ppm}).

Finally, objects are considered as member stars according to the photometric membership as well as kinematic one.
Based on the CMD and PMs analysis, we identify 959 member stars in total within the $40\arcmin \times 40\arcmin$ field.
Among the 42 emission-line candidates, 23 objects are eligible for our criteria and suggested to be member stars.

\begin{figure}
\epsscale{1}
\plotone{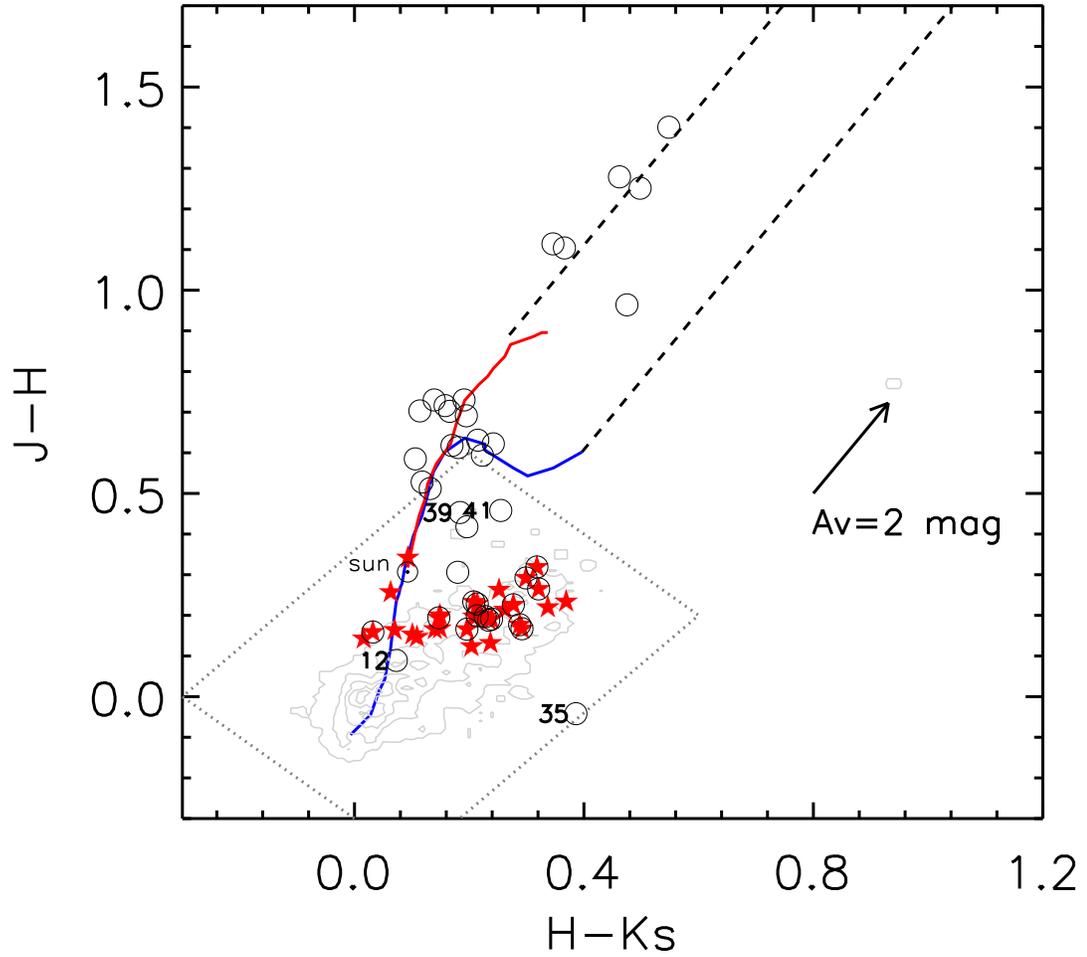}
\caption{2MASS color--color diagram in the cluster region. The red and blue curves show the giant and dwarf loci \citep{bnb88} converted to the 2MASS
system. The arrow represents the reddening direction \citep{rei85} for typical Galactic interstellar extinction ($R_V=3.1$), and the dashed lines encompass
the region of reddened giants and dwarfs. The grey contours demonstrate known Be stars distribution, and we define a dashed box to include all possible Be stars.
Symbols are same as Figure~\ref{Ha}.}
\label{tcd}
\end{figure}

To further select possible Be stars from these 23 emission-line candidates, we defined a region in the $J-H$ vs. $H-K_{s}$ color-color diagram, which could cover 99\%
of Be stars collected from literatures \citep{zha05}. As shown in Figure~\ref{tcd}, grey contours represent over one thousand Be stars.
Assuming that most Be stars have similar infrared colors, we thus selected possible Be stars inside grey-dashed region in the color-color diagram. 
Finally, 19 candidates are selected as Be stars in NGC\,663. Among the 19 candidates, we have confirmed that 15 objects are previously reported Be stars. 
Therefore, we discover 4 Be stars (ID number:~12, 35, 39, and 41) in NGC\,663.

\section{Results and Discussion}
\subsection{Known Be Stars}
\citet{pig01} listed the compilation of 25 known Be stars in NGC\,663 from previous studies. They further discovered 4 new Be stars using H$\alpha$ 
photometry with a 60-cm telescope, and presented a list of 29 Be stars in the 30\arcmin~$\times$ 30\arcmin~field. 
\citet{mathew11} found 2 more new Be stars with a spectroscopic observation and increased the number to 31.
We list the properties of these 31 known Be stars in Table~2. The first column provides the names of known Be stars. Columns 2 and 3 provide coordinates in degree. 
Columns 4 and 5 list the PM measurements in right ascension and declination adopted from \citet{kha12}. Column 6 show errors of the PM measurements. 
Subsequent columns, 7, 8, and 9 list the 2MASS $K_{s}$, $J-H$ and $H-K_{s}$ values.
Column 10 show the H$\alpha$ equivalent widths (EW(H$\alpha$)) adopted from \citet{mathew11}. 
We use the averaged EW(H$\alpha$) if the Be stars have multiple observations. 
Column 11 indicates the spectral type adopted from \citet{mathew11}. The final column shows the notes of the stars.

\begin{figure}
\epsscale{1.0}
\plotone{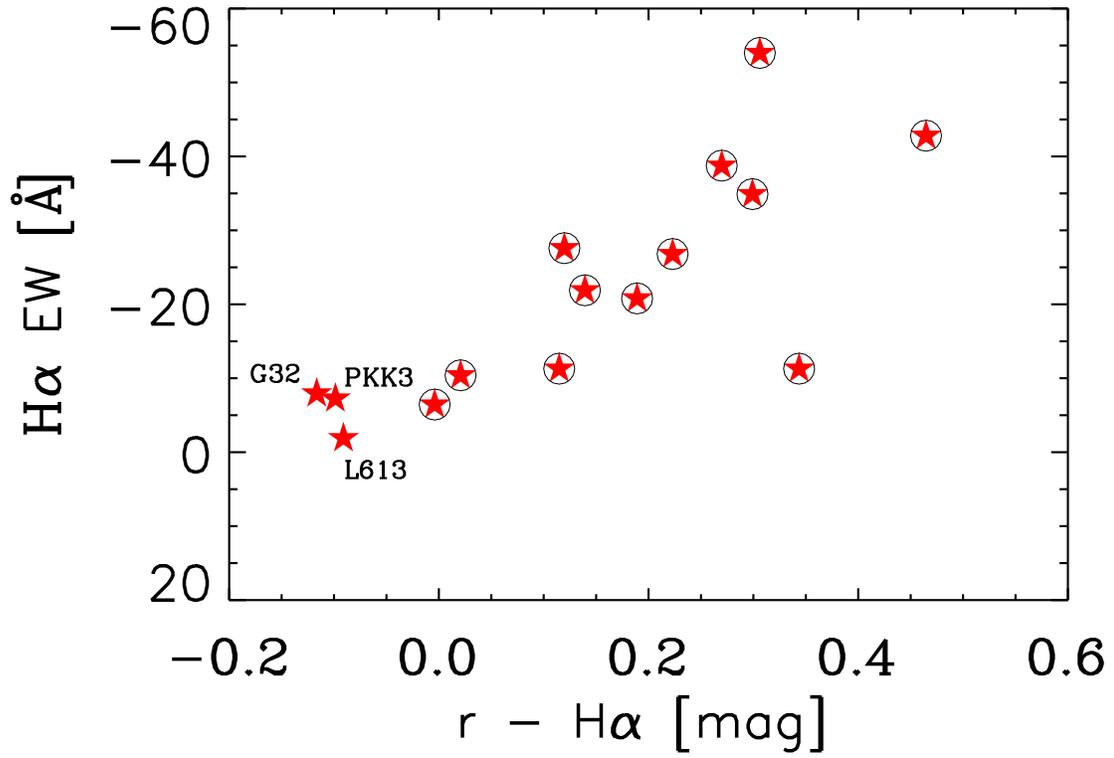}
\caption{EW(H$\alpha$) versus r{$-$}H$\alpha$. Symbols are same as Figure~\ref{Ha}. There is a correlation between the EW(H$\alpha$) and r${-}$H$\alpha$ values.
Three known Be stars (PKK3, L613, and G32) with weak H$\alpha$ strength ($>$ $-$10\AA) are not selected by our method.}
\label{EW}
\end{figure}

Among 31 known Be stars, we re-identify 15 objects and misidentify 6 objects. The rest 10 objects are discounted because of
photometric problems, including saturation, blending of nearby stars, and contamination of the CCD gap.
The misidentification of 6 objects might be caused by the variability; we find that two of the misidentified objects GG99 and L613 are suggested to show H$\alpha$ variability \citep*{Sanduleak90,mathew11}.
Another reason of the misidentification might be due to weak emissions. As shown in Figure 2, the 4 misidentified objects (G32, PKK1, PKK2, and PKK3) show low r${-}$H$\alpha$ values, which are similar to the mean values. This result indicates that their H$\alpha$ strength could be too weak for firm identification. Figure~\ref{EW} shows the correlation between the EW(H$\alpha$) and r${-}$H$\alpha$ values,
suggesting the EW(H$\alpha$) of the 4 misidentified objects are weaker than $-$10\AA. Therefore, the misidentification of the 4 objects could be due to the weak emissions.
If we select emission-line candidates with a lower threshold r$-$H$\alpha$ $>$ 1$\sigma_{p}$ to include objects with weak H$\alpha$, we then have additional 124 emission-line candidates with lower confidence levels
while only two more known Be stars (G32 and PKK3) can be identified as emission-line stars. Thus, our detection limit for emission-line candidate is about EW(H$\alpha$) $=$ $-$10\AA~(r$-$H$\alpha$ $>$ 2$\sigma_{p}$).

\subsection{New Be Candidates}
We discover four possible new Be stars (ID number:~12, 35, 39, and 41) with r${-}$H$\alpha$ $>$~0.12.  
The EW(H$\alpha$) versus r${-}$H$\alpha$ correlation (Figure~\ref{EW}) shows these newly discovered Be stars to have EW(H$\alpha$) stronger than~$-$30~\AA,
thus the undetection of the 4 new Be stars in previous studies should not be caused by weak emissions.
In addition to the emission-line variability, the undetection of the 4 new Be stars in previous studies might be due to their positions.
Previous survey of NGC\,663 covered only less the central 30\arcmin~$\times$ 30\arcmin~field, whereas two of the four new Be stars are located outside region.
Besides, by the same diagnosis, we find that one previously claimed Be star SAN\,28 should not be the member star due to its inconsistent PM ($\mu_{\alpha}=-16.07\pm3.95, \mu_{\delta}=-34.09\pm3.95$~mas~yr$^{-1}$).
We thus conclude that the number of Be stars in NGC\,663 is 34.

\subsection{Be Star Fraction}
\citet{mat08} discovered 22 Be candidates among a total of 486 B-type stars in NGC\,663.
They further compare the ratio of Be stars to B-type stars [N(Be)/N(Be+B)] of NGC\,663 with that of two rich clusters NGC\,7419 and NGC\,2345. 
Since NGC\,663, NGC\,7419 and NGC~2345 are all rich, moderately young open clusters with ages between 31~Myrs and 79~Myrs \citep{kha13}, 
NGC\,663 could have similar Be star fraction with that of NGC\,7419 and NGC\,2345.
Interestingly, they showed the [N(Be)/N(Be+B)] fraction of NGC\,663 is 4.5\%, which is lower than that of NGC\,7419 (11\%) and NGC\,2345 (26\%).
In spite of that our result increases the [N(Be)/N(Be+B)] fraction in NGC\,663 from 4.5\% to 6.8\%, the fraction is still the lowest among these three clusters.

It is notable that our procedure of membership identification is quite different from that of \citet{mat08}. 
Since they did not consider kinematic information to identify memberships, we perform the same method (see Sec.~2.1.3) for NGC\,7419 and NGC\,2345 
to identify membership, and make a consistent comparison among these 3 clusters.
Consequently, the ratio of Be stars to member stars [N(Be)/N(members)] of NGC\,663, NGC\,2345 and NGC\,7419 are 3.5\% (34/959), 6.8\% (11/160), and 10.1\% (22/217).
Once again NGC\,663 shows the lowest ratio among these 3 clusters.
One possibility explains the low Be star fraction in NGC\,663 is that the circumstellar material might be swept away by stellar winds from massive stars or supernova explosions. 
Since NGC\,663 could be a part of stellar association Cas\,OB8, stellar winds from nearby massive stars or supernova events might remove the  circumstellar material.
Moreover, \citet{pandey05} reported that massive stars tend to lie in the centre of NGC\,663, and suggested that the mass segregation is primordial, e.g. due to star formation process. 
When higher-mass stars preferentially locate towards in the centre of NGC\,663, the circumstellar material might be swept in the early stage of star-forming process, and thus cause the low Be star fraction.

\subsection{Spectral Type}
The spectral types of some Be stars in NGC\,663 are present by \citet{mathew11}. It might be able to classify the stars based on the relative brightness of the stars with and without spectra.
According to known spectra, the stars with r-band instrumental magnitudes brighter than 8.5~mag are classified as B0{--}B2V, and the ones with magnitudes between 9~and 11~magnitudes are belong to B5{--}B7 stars. 
Thus, we could classify the Be stars with magnitudes between 8.5~and 9~magnitudes as B3{--}B4 stars. Those candidates with r-band magnitudes fainter than 11~mag might be late-type B or early-type A stars, e.g., B8{--}A0 stars. Consequently, we have 21 Be stars in B0{--}B2 type including saturated stars, which should be brighter than 8.5~mag. We also have 2 Be stars in B3{--}B4 type, 7 ones in B5{--}B7 type, and 3 ones 
in B8{--}A0 type. Our result shows similar distribution of spectral type to that of previous studies; \citet{mer82} investigated the distribution of Be stars as a function of spectral types and found that a maxima at type of B1{--}B2 and B7{--}B8 while \citet{mat08} showed bimodal peaks in B1{--}B2 and B5{--}B7 for open clusters. Among four newly discovered Be stars, one is classified as a B3{--}B4 star and three should
be B8{--}A0 stars. Before we discover 3 Be candidates belong to B8{--}A0 type, the fraction of Be stars with spectral type later than B5 is 21\% (7/33) in NGC\,663. 
Our results increase the fraction to 30\% (10/33), and make it to be consistent with other open clusters \citep{mat08}.

\subsection{New Emission-Line Candidates}
We have identified 42 emission-line stars within a $40\arcmin \times 40\arcmin$ field of NGC\,663; 15 stars are known Be stars and 4 stars are
newly discovered. Among the remaining 23 candidates, 2 candidates could be emission-line dwarf stars and 6 candidates could be giants in NGC\,663. 
For the rest 15 candidates, we have also identified 2 stars as Be stars and 13 objects as emission-line dwarfs in the field that are not associated with the cluster.
After cross-matching with the SIMBAD database, these 23 emission-line stars are presented for the first time.

\section{SUMMARY}
We apply the H$\alpha$ imaging photometry to identify 42 emission-line candidates in NGC\,663. Four newly discovered Be stars have been identified as members with the CMD and PMs.
We also discover that one known Be star SAN28 should not be a cluster member due to its inconsistent PMs. 
Our results show that the number of Be stars in NGC\,663 is 34, and the ratio of Be stars to member stars [N(Be)/N(members)] is 3.5\%. 
The low fraction of Be stars in NGC\,663 might be the consequence of the effects of mass segregation and stellar winds, or supernova explosions.  
Our results also reveal bimodal peaks of spectral type B0--B2 and B5--B7 for the Be population, which is consistent with the statistics in most star clusters. 
Finally, we discover 23 emission-line stars of non-member Be stars, dwarfs, and giants. 
Our results suggest that the H$\alpha$ photometry of the PTF data is very useful to search for emission-line stars.
More Be stars in nearby clusters will be further searched and investigated to study the evolutionary scenario of the Be phenomena. 

\acknowledgments
We thank the referee for constructive comments. We also thank C.~K.~Chang, M.~Kasiliwal, and H.~C.~Chen for useful discussions and information. 
This work is supported in part by the National Science Council of Taiwan under grants NSC 101-2119-M-008-007-MY3 (W.-H.I.), NSC 102-2119-M-008-001 (W.-P.C.), and NSC101-2112-M-008-017-MY3 (C.-C.N.). This publication makes use of data products from the Two Micron All Sky Survey, which is a joint project of the University of Massachusetts and the Infrared Processing and Analysis Center/California Institute of Technology, funded by the National Aeronautics and Space Administration and the National Science Foundation.

% spatial distribution

% 2MASS color-magnitude diagram

% stellar proper motions

% 2MASS color-color diagram

\clearpage
\begin{deluxetable}{lrrrrrrrrrrrr}
\rotate
\tabletypesize{\small}
\tablecolumns{12} \tablewidth{0pt}\tablecaption{Emission-Line Candidates}
\tablehead{
\colhead{ID} &
\colhead{R.A.} &
\colhead{Dec.} &
\colhead{$\mu_{\alpha}$} &
\colhead{$\mu_{\delta}$} &
\colhead{epm} &
\colhead{$K_{s}$} &
\colhead{$J{-}H$} &
\colhead{$H{-}K_{s}$} &
\colhead{Name} &
\colhead{r} &
\colhead{$r{-}H\alpha$} &
\colhead{SIMBAD}  \\
\colhead{} &
\colhead{deg} &
\colhead{deg} &
\colhead{mas/yr} &
\colhead{mas/yr} &
\colhead{mas/yr} &
\colhead{mag} &
\colhead{mag} &
\colhead{mag} &
\colhead{} &
\colhead{mag} &
\colhead{mag} &
\colhead{}
}
\startdata
1  &26.483740  &61.212574  &$-$1.35  &4.68  &1.50  &9.675  &0.266  &0.321  &GC97  &7.715  &0.120 &Be\\
2  &26.443295  &61.155804  &$-$1.79  &$-$1.36  &2.00  &9.895  &0.198  &0.228  &GG95  &7.647  &0.270 &Be\\
3  &26.326494  &60.969196  &$-$4.48  &2.90  &2.00  &9.751  &0.226  &0.278  &GG90  &7.415  &0.027 &Be\\
4  &26.750830  &61.356567  &$-$2.56  &$-$1.37  &2.00  &9.486  &0.200  &0.213  &GG104  &7.341  &0.032 &HXB\\
5  &26.861519  &61.145599  &$-$3.74  &$-$6.38  &2.00  &10.237  &0.158  &0.033  &GG108  &7.476  &0.344 &Be\\
6  &26.739870  &61.027885  &$-$5.86  &0.82  &2.00  &9.937  &0.189  &0.233  &GG103  &7.655  &$-$0.022 &Be\\
7  &26.415096  &61.216434  &$-$2.40  &3.04  &2.00  &10.494  &0.191  &0.238  &GG94  &8.107  &0.139 &Be\\
8  &26.407560  &61.133095  &0.63  &$-$2.93  &2.00  &10.644  &0.167  &0.291  &GG93  &7.920  &0.299 &Be\\
9  &26.765589  &61.292236  &$-$3.92  &0.93  &2.00  &10.534  &0.234  &0.207  &D01$-$034  &8.221  &0.115 &Be\\
10  &26.627619  &61.241455  &$-$7.84  &$-$1.38  &2.00  &10.463  &0.165  &0.196  &VES620  &8.046  &0.189 &Be\\
11  &26.202705  &61.215088  &1.04  &$-$2.00  &5.00  &7.284  &1.104  &0.367  &---  &8.505  &$-$0.044 &---\\
12  &26.108356  &60.895775  &2.89  &1.26  &2.00  &11.642  &0.090  &0.073  &---  &8.665  &0.229  &---\\
13  &26.914015  &61.305721  &$-$3.06  &$-$0.88  &2.79  &10.144  &0.292  &0.298  &GG109  &8.295  &0.465 &Be\\
14  &26.648338  &61.227543  &$-$4.11  &2.39  &2.00  &10.126  &0.318  &0.318  &GG101  &8.284  &0.306 &Be\\
15$^{*}$  &26.320665  &61.123646  &$-$98.90  &$-$8.50  &12.90  &9.852  &0.716  &0.158  &---  &8.823  &$-$0.008 &iC\\
16  &26.068714  &61.076244  &$-$0.98  &2.74  &5.00  &7.892  &1.115  &0.345  &---  &9.185  &$-$0.038 &---\\
17$^{*}$  &26.444133  &60.972193  &357.10  &53.10  &5.00  &9.565  &0.571  &0.240  &---  &9.068  &0.121 &PM\\
18  &26.558376  &61.228851  &$-$8.65  &0.34  &2.00  &11.168  &0.177  &0.288  &BG114  &9.172  &0.021 &Be\\
19  &26.748318  &61.208199  &$-$5.21  &2.70  &2.59  &11.590  &0.227  &0.215  &VES624  &9.466  &0.223 &Be\\
20  &26.431408  &61.415092  &$-$3.61  &$-$3.60  &5.00  &9.163  &0.965  &0.474  &---  &9.943  &0.047 &---\\
21  &26.230267  &61.005119  &9.26  &$-$11.48  &3.90  &11.725  &0.585  &0.106  &---  &9.941  &$-$0.059 &---\\
22  &26.943848  &61.370613  &29.70  &$-$52.68  &3.90  &11.112  &0.731  &0.139  &---  &10.141  &$-$0.053 &---\\
23  &26.036928  &61.300438  &$-$1.89  &$-$0.89  &5.00  &7.337  &1.400  &0.549  &---  &10.409  &0.083  &---\\
24  &26.223892  &61.233059  &$-$7.30  &$-$15.64  &3.90  &11.527  &0.701  &0.167  &---  &10.376  &$-$0.032 &iC\\
25  &27.009970  &61.320004  &28.46  &$-$15.60  &3.90  &11.623  &0.632  &0.215  &---  &10.747  &$-$0.040 &---\\
26  &26.823006  &61.221569  &$-$2.82  &0.51  &3.90  &12.619  &0.194  &0.148  &PKK4  &10.308  &$-$0.004 &Be\\
27  &27.185598  &60.894218  &0.07  &$-$2.06  &5.00  &7.355  &1.280  &0.462  &---  &10.430  &0.030 &---\\
28  &26.424812  &61.442692  &$-$3.41  &$-$1.26  &3.90  &12.887  &0.511  &0.133  &---  &11.000  &$-$0.034 &---\\
29  &27.068419  &61.100658  &$-$2.23  &$-$4.46  &5.00  &8.112  &1.252  &0.497  &---  &11.115  &0.078 &---\\
30  &26.023653  &61.306664  &24.85  &$-$19.31  &4.00  &11.954  &0.691  &0.195  &---  &11.444  &0.146 &---\\
31  &26.088741  &61.198486  &1.61  &3.07  &4.00  &12.998  &0.618  &0.170  &---  &11.702  &0.136 &---\\
32$^{*}$  &26.698837  &61.124264  &$-$233.60  &$-$5.40  &15.30  &11.920  &0.594  &0.223  &---  &11.263  &0.021 &iC\\
33  &26.466228  &61.297523  &$-$12.96  &$-$15.77  &3.90  &13.717  &0.418  &0.196  &---  &12.167  &0.079 &iC\\
34  &26.037413  &61.119137  &52.73  &$-$19.68  &12.07  &13.852  &0.306  &0.180  &---  &11.892  &0.067 &---\\
35$^{*}$  &26.322468  &61.047340  &7.40  &7.20  &4.00  &14.071  &$-$0.042  &0.386  &---  &12.163  &0.119 &---\\
36  &26.706167  &61.425907  &9.88  &$-$23.59  &3.90  &12.303  &0.623  &0.242  &---  &11.875  &0.088 &---\\
37  &26.877998  &61.107220  &$-$21.54  &$-$10.89  &4.65  &12.812  &0.729  &0.191  &---  &11.924  &0.073 &---\\
38  &26.213600  &61.322948  &26.75  &$-$37.80  &15.86  &13.727  &0.613  &0.182  &---  &12.575  &0.242 &---\\
39  &26.347054  &61.265411  &4.05  &7.71  &4.00  &14.227  &0.453  &0.184  &---  &12.732  &0.256 &---\\
40  &26.411844  &60.933365  &45.93  &$-$61.68  &3.90  &13.666  &0.703  &0.113  &---  &12.640  &0.334 &---\\
41  &26.214262  &60.938358  &$-$14.28  &$-$6.80  &4.00  &14.206  &0.457  &0.255  &---  &12.544  &0.161 &---\\
42  &26.976549  &60.914070  &62.77  &$-$33.19  &5.79  &13.935  &0.528  &0.118  &---  &12.686  &0.158 &---\\
\enddata
\tablecomments{Column 1: ID numbers of candidates. Numbers with asterisks: No matched star in \citet{kha12}, and we adopt PMs from PPMXL.
Column 2: RA in degree (J2000). Column 3: Dec in degree (J2000). Column 4: Proper motion of RA.
Column 5: Proper motion of Dec. Column 6: PM uncertainties. 
Column 7: $K_{s}$-band magnitudes adopted from 2MASS. Column 8: $J{-}H$. Column 9: $H{-}K_{s}$. Column 10: Known name.
Column 11: r$-$band instrumental magnitude. Column 12: r$-$H$\alpha$ magnitudes.
Column 13: Object classification adopted from SIMBAD; Be: Be stars; HXB: high mass X-ray binary; iC: star in cluster; PM: high proper-motion star.}
\end{deluxetable}

\clearpage
\begin{deluxetable}{lrrrrrrrrrrr}
\rotate
\tabletypesize{\small}
\tablecolumns{12} \tablewidth{0pt}\tablecaption{Properties of Known Be stars in NGC\,663}
\tablehead{
\colhead{Name} &
\colhead{R.A.} &
\colhead{Dec.} &
\colhead{$\mu_{\alpha}$} &
\colhead{$\mu_{\delta}$} &
\colhead{epm} &
\colhead{Ks} &
\colhead{$J{-}H$} &
\colhead{$H{-}Ks$} &
\colhead{EW (H$\alpha$)} &
\colhead{Spectral type} &
\colhead{Note}\\
\colhead{} &
\colhead{deg} &
\colhead{deg} &
\colhead{mas/yr} &
\colhead{mas/yr} &
\colhead{mas/yr} &
\colhead{mag} &
\colhead{mag} &
\colhead{mag} &
\colhead{\AA} &
\colhead{} &
\colhead{}
}
\startdata
 %  name         ranew      denew                pmra     pmde    era       kmag       j-h          h-k                     EW              Sp type       Note
MWC\,428  &26.325104  &61.115692   &$-$5.05    &1.57   &2.00   &8.602    &0.196   &0.209         &---               &---             &Saturated\\
GG\,102     &26.648041  &61.263290    &$-$4.64   &$-$0.74  &2.02   &8.880    &0.132   &0.236     &---               &---            &Saturated\\
MWC\,700  &26.612782  &61.167202    &$-$5.76   &1.39   &2.00   &8.987    &0.144   &0.015         &---               &---            &Saturated\\
VES\,619    &26.611837  &61.128258    &$-$5.26   &2.14   &2.00   &9.229    &0.221   &0.337         &---               &---            &Saturated\\
BG\,15        &26.615303  &61.207062    &$-$6.65   &4.92   &2.00   &9.251    &0.201   &0.148          &---               &---            &Saturated\\
MWC\,698  &26.497110  &61.212685    &$-$3.69   &$-$6.16  &2.89   &9.425    &0.148   &0.109      &---               &---            &Saturated\\
VES\,616    &26.525537  &61.227573    &$-$2.48   &$-$5.76  &2.00   &9.645    &0.234   &0.368      &---               &---            &Saturated\\
GG\,110      &27.096041  &61.264717    &$-$2.96   &$-$3.82  &2.00   &9.704    &0.168   &0.149      &---               &---            &Saturated\\
GG\,104      &26.750830  &61.356567    &$-$2.56   &$-$1.37  &2.00   &9.899    &0.200   &0.213     &---               &---             &---\\
GG\,90        &26.326494  &60.969196    &$-$4.48   &2.90   &2.00   &10.255  &0.226   &0.278         &---               &---             &---\\
GC\,97        &26.483740  &61.212574    &$-$1.35   &4.68   &1.50   &10.262  &0.266   &0.321         &$-$27.6       &B2V         &---\\
GG\,95        &26.443295  &61.155804    &$-$1.79   &$-$1.36  &2.00   &10.321  &0.198   &0.228     &$-$38.8       &B1V         &---\\
GG\,103      &26.739870  &61.027885    &$-$5.86   &0.82   &2.00   &10.359  &0.189   &0.233         &---               &---             &---\\
GG\,98        &26.584209  &61.239319    &$-$6.07   &0.02   &2.00   &10.402  &0.213   &0.262         &---               &---             &Blended\\
GG\,108      &26.861519  &61.145599    &$-$3.74   &$-$6.38  &2.00   &10.428  &0.158   &0.033     &$-$11.3       &B0--B1V   &---\\
GG\,109      &26.914015  &61.305721    &$-$3.06   &$-$0.88  &2.79   &10.734  &0.292   &0.298     &$-$42.8       &B1V           &---\\
GG\,101      &26.648338  &61.227543    &$-$4.11   &2.39   &2.00   &10.762   &0.318   &0.318         &$-$54.0       &B1V          &---\\
VES\,620    &26.627619  &61.241455    &$-$7.84    &$-$1.38  &2.00  &10.824   &0.165   &0.196     &$-$20.8       &B1V          &---\\
GG\,99       &26.612082   &61.236439    &$-$5.19    &1.75   &2.00  &10.880   &0.153   &0.101         &---               &---         &---\\
GG\,94       &26.415096   &61.216434    &$-$2.40    &3.04   &2.00  &10.923   &0.191   &0.238         &$-$21.9       &B2V         &---\\
D01$-$034     &26.765589   &61.292236    &$-$3.92    &0.93   &2.00  &10.975   &0.234   &0.207     &$-$11.3       &---         &---\\
GG\,93       &26.407560   &61.133095    &0.63     &$-$2.93  &2.00  &11.102   &0.167   &0.291          &$-$34.9       &B2V         &---\\
G\,32          &26.619238   &61.230675    &$-$4.15    &2.47   &2.00  &11.338   &0.125   &0.203          &$-$7.9         &B3V         &---\\
BG\,114      &26.558376   &61.228851    &$-$8.65    &0.34   &2.00  &11.633   &0.177   &0.288          &$-$10.4       &B5V         &---\\
VES\,624    &26.748318   &61.208199    &$-$5.21    &2.70   &2.59  &12.032   &0.227  &0.215           &$-$26.8      &B5V         &---\\
L\,613         &26.645269   &61.107712    &$-$0.09    &$-$0.27  &2.59  &12.126   &0.163  &0.071       &$-$1.9        &B5V         &---\\
SAN\,28      &26.508623  &61.250595    &$-$16.07  &$-$34.09  &3.95 &12.388   &0.262  &0.253       &$-$42.1      &B5V         &CCD gap\\
PKK\,1        &26.298687   &61.173038    &11.65   &$-$7.07  &3.90  &11.864   &0.342   &0.092          &---                 &---          &---\\
PKK\,2        &26.449562   &61.273327    &$-$0.43    &$-$2.75  &3.90  &12.060   &0.257  &0.064       &---               &---            &---\\
PKK\,3        &26.601690  &61.177017    &$-$1.70    &$-$1.61   &2.59  &12.225   &0.166  &0.141       &$-$7.3        &B5--B7V         &---\\
PKK\,4        &26.823006  &61.221569    &$-$2.82    &0.51    &3.90  &12.961   &0.194  &0.148           &$-$6.5        &B5--B7V         &---\\

\enddata
\tablecomments{Column 1: Name of known Be stars. Column 2: R.A. in degree (J2000). Column 3: Dec. in Degree (J2000).
Column 4: Proper motions of R.A. Column 5: Proper motions of Dec.  Column 6: Uncertainties of proper motions.
Column 7: $K_{s}$-band magnitudes adopted from 2MASS. Column 8: $J{-}H$ magnitudes. Column 9: $H{-}K_{s}$ magnitudes. 
Column 10:  H$\alpha$ equivalent widths adopted from \citet{mathew11}. If the objects have more than one observation at different epoches,
we take the averaged equivalent widths. Column 11: Spectral type adopted from \citet{mathew11}. Column 12: Notes of objects.}
\end{deluxetable}

\end{CJK*}

\end{document}